\documentclass{INTERSPEECH2023}

\interspeechcameraready 

\usepackage{float}
\usepackage{algorithm}
\usepackage{algpseudocode}
\usepackage[fontsize=9pt]{fontsize}
\usepackage{bold-extra}
\usepackage[T1]{fontenc}
\usepackage[subtle]{savetrees}
\usepackage{amssymb}
\usepackage{pifont}
\usepackage{cite}
\usepackage{cleveref}
\usepackage{ctable}
\usepackage{multirow}
\usepackage{lipsum}
\usepackage{caption}
\usepackage{subcaption}
\usepackage{booktabs}
\usepackage{hyperref}

\newcommand{\newpara}[1]{\vspace{3pt}\noindent\textbf{#1}}

\newcommand{\cmark}{\ding{51}}%
\newcommand{\xmark}{\ding{55}}%
\newcommand\blfootnote[1]{%
  \begingroup
  \renewcommand\thefootnote{}\footnote{#1}%
  \addtocounter{footnote}{-1}%
  \endgroup
}

\title{Disentangled Representation Learning for Multilingual Speaker Recognition}
\name{Kihyun Nam$^{1\dag}$, Youkyum Kim$^{1\dag}$, Jaesung Huh$^2$, Hee-Soo Heo$^3$, Jee-weon Jung$^4$, Joon Son Chung$^1$}

\address{
  $^1$Korea Advanced Institute of Science and Technology, South Korea\\
  $^2$University of Oxford, United Kingdom \\
  $^3$Naver Corporation, South Korea \\
  $^4$Carnegie Mellon University, USA}
\email{joonsc@kaist.ac.kr}

\begin{document}
\maketitle

\blfootnote{\hspace{-4pt}$^{\dag}$These authors contributed equally to this work.}
 
\begin{abstract}

The goal of this paper is to learn robust speaker representation for bilingual speaking scenario. The majority of the world's population speak at least two languages; however, most speaker recognition systems fail to recognise the same speaker when speaking in different languages. 

Popular speaker recognition evaluation sets do not consider the bilingual scenario, making it difficult to analyse the effect of bilingual speakers on speaker recognition performance. In this paper, we publish a large-scale evaluation set named VoxCeleb1-B derived from VoxCeleb that considers bilingual scenarios.

We introduce an effective disentanglement learning strategy that combines adversarial and metric learning-based methods. This approach addresses the bilingual situation by disentangling language-related information from speaker representation while ensuring stable speaker representation learning. Our language-disentangled learning method only uses language pseudo-labels without manual information.

\end{abstract}
 \section{Introduction}
\label{sec:intro}

An estimated 60 to 75 percent of the world's population speaks at least two languages~\cite{vince2016amazing}. While somebody is speaking in a foreign language, it has been observed that the person's voice sounds different from when speaking in their mother tongue~\cite{lee2017bilingual}. 
With recent trends in globalisation, it has become easier to encounter multilingual scenarios. Therefore, the focus on multilingual speaker recognition has become more important~\cite{cieri2007resources, akbacak2007language, ferrer2011promoting, reynolds20172016, matejka2017analysis}. 

While the performance of speaker recognition systems has improved significantly due to recent advances in deep learning \cite{he2016deep, snyder2018x, jung2018complete, ravanelli2018speaker, snyder2019speaker, snell2017prototypical, khosla2020supervised, kang2022augmentation} and the availability of large-scale datasets~\cite{nagrani2020voxceleb, chung18b_interspeech}, the state-of-the-art systems fail easily under the language mismatch condition. 
The popular speaker recognition evaluation sets do not consider bilingual scenarios, making it difficult to analyse their effect on speaker recognition performance. 
There are a few evaluation datasets that consider bilingual scenarios; however, they are collected from controlled environments like phone-call platform~\cite{cieri2007resources}  or contain only limited languages~\cite{reynolds20172016}.
The recent VoxCeleb Speaker Recognition Challenge (VoxSRC)~\cite{brown2022voxsrc} contains some bilingual speakers; however, their evaluation datasets remain private. 
Hence, to the best of our knowledge, there is no large-scale public evaluation set that takes bilingual speakers into account.

To this end, we publish a large-scale bilingual evaluation set derived from VoxCeleb1~\cite{nagrani2020voxceleb}, focusing on bilingual speaking problems. We call this test protocol \texttt{VoxCeleb1-B}\footnote{The data is available from: \href{https://mm.kaist.ac.kr/projects/voxceleb1-b/}{{https://mm.kaist.ac.kr/projects/voxceleb1-b/}}}. To increase the scale and the diversity compared to the VoxSRC challenge test set~\cite{brown2022voxsrc}, we expand the number of bilingual trials and the number of languages, resulting in a total of 808,574 trials and 15 languages. Moreover, for the first time, we release the manually annotated language labels of VoxCeleb1. 
More details of VoxCeleb1-B and the language labels are given in Section \ref{sec:VoxCeleb1-B}.

Previous literature finds that a speaker's identity information is intertwined with various factors including accent~\cite{raj2019probing}, gender~\cite{luu21_interspeech,luu2022investigating}, age~\cite{luu2022investigating}, nationality~\cite{luu2022investigating}, emotion~\cite{williams2019disentangling, pappagari2020x}, and spoken language~\cite{maiti2020generating}. Using the proposed evaluation protocol, we observe that the existing speaker recognition models do not generalise well to bilingual speakers. We suppose that the mismatched prosodic characteristics from bilingual speakers' different languages significantly affect the performance of the speaker recognition models.

To resolve the language-dependent problem, traditional methods on multilingual speaker recognition have mostly utilised combination of probabilistic linear discriminant analysis and scoring functions based on a standard backbone system such as the i-vectors~\cite{ferrer2011promoting, misra2014spoken, matejka2017analysis}. However, these methods do not ensure language-invariant speaker representations. Other studies~\cite{meng2019adversarial, xia2019cross, xin2021cross, kang2022augmentation, kang2020disentangled, kwon2020intra, tong22_odyssey, yi2022disentangled, mun2022disentangled} have proposed two types of disentangled representation learning methods, namely \textit{adversarial learning-based method} and \textit{metric learning-based method}, which isolate nuisance attributes from the speaker representation. Adversarial learning-based method disturbs convergence of non-speaker discriminator, while metric learning-based method minimises distance or similarity between speaker-relevant and non-speaker representations. For adversarial learning-based method, some studies~\cite{meng2019adversarial, xia2019cross, xin2021cross, kang2022augmentation} utilise the gradient reversal layer (GRL). Although GRL has shown performance improvement in disentanglement of the target information, we find through our experiments that it frequently causes unstable training and is sensitive to hyperparameters. On the other hand, some studies \cite{kang2020disentangled, tong22_odyssey, yi2022disentangled, mun2022disentangled} propose metric learning-based methods to minimise correlation between speaker representation and non-speaker representations. \cite{kang2020disentangled} utilises mean absolute Pearson’s correlation (MAPC) minimisation and \cite{tong22_odyssey} uses cosine similarity (COS) minimisation. \cite{yi2022disentangled, mun2022disentangled} employ mutual information minimisation. However, since \cite{kang2020disentangled, yi2022disentangled, mun2022disentangled} perform domain adaptation for different domains, there is no guarantee that they will perform well in the goal of this work, namely \textit{intra-domain disentangled representation learning.} We evaluate the existing methods and our method on the evaluation set which reflects real-world bilingual scenario unlike \cite{tong22_odyssey} which conducts experiments on a simulated dataset.

\begin{table*}[t!]
\centering
\caption{Statistics of the VoxCeleb1 test sets, VoxSRC validation sets and VoxCeleb1-B. {\bf Pos.}: \# of positive trials; {\bf Neg.}: \# of negative trials; {\bf cl.}: Cleaned version; {\bf Cross-lingual}: Whether the test set is constructed in consideration of bilingual scenario.}
\resizebox{1.0\linewidth}{!}{
\begin{tabular}{@{\extracolsep{8pt}}lcccccc@{}}
\toprule
{\bf Test set} & {\bf VoxCeleb1 cl.} & {\bf VoxCeleb1-E cl.} & {\bf VoxCeleb1-H cl.} & {\bf VoxSRC 2020 Val} & {\bf VoxSRC 2021 Val} & {\bf VoxCeleb1-B}  \\
\toprule
\# of trials & 37,611 & 579,818 & 550,894 & 263,486 & 60,000 & 808,574 \\
(Pos. / Neg.) & (18,802 / 18,809) & (289,921 / 289,897) & (275,488 / 275,406) & (131,743 / 131,743) & (29,969 / 30,031)  & (404,287 / 404,287) \\
\toprule
Cross-lingual & \xmark & \xmark & \xmark & \xmark & \cmark & \cmark\\
\bottomrule
\end{tabular}}
\label{tab:testsets}
\end{table*}

\begin{figure}[!t]
\centering
\vspace{1mm}
{
    \begin{subfigure}[b]{\linewidth}
         \centering
         \includegraphics[width=\textwidth]{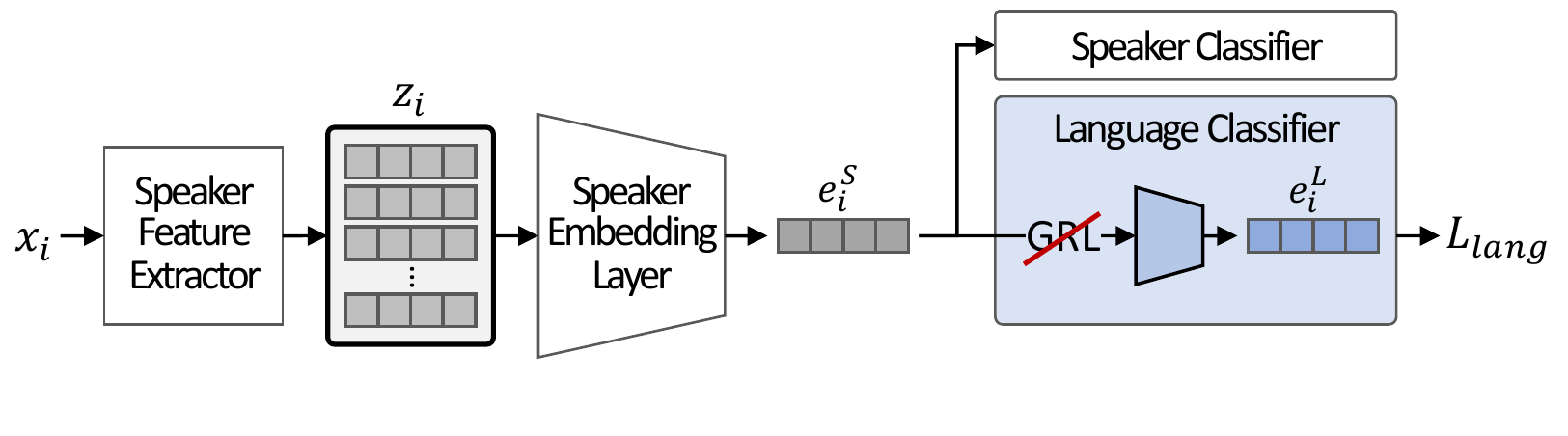}
         \vspace{-7mm}
         \caption{Language discriminator training}
         \label{fig:discriminator_training}
     \end{subfigure}
     \begin{subfigure}[b]{\linewidth}
         \centering
         \includegraphics[width=\textwidth]{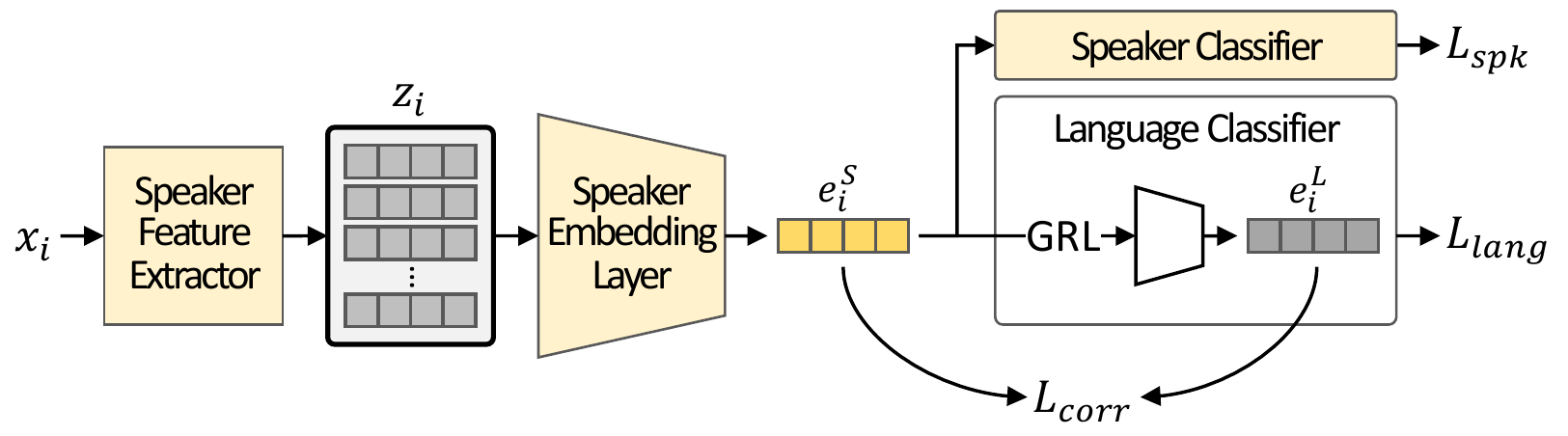}
         \vspace{-5mm}
         \caption{Speaker embedding training}
         \label{fig:representation_training}
     \end{subfigure}
}
\vspace{-4mm}
\caption{Overview of the training strategy. The coloured parts of the network are updated during each training procedure. Note that the Gradient Reversal Layer (GRL) is only activated during speaker embedding training procedure. \bm{$x_i$}: input mel-spectrogram; \bm{$z_i$}: frame-level embeddings; \bm{${e_i}^S$}: speaker embedding vector; \bm{${e_i}^L$}: language feature vector.}
\vspace{-3mm}
\label{fig:training_framework}
\end{figure}

In this work, we propose an effective disentangled representation learning, which removes the language-dependent information in the speaker representation. The proposed learning strategy combines GRL and MAPC minimisation objective, which overcomes unstable learning and effectively learns language-disentangled speaker representation. The neural network consists of a main speaker recognition model and a spoken language recognition model. During training, language-disentangled learning leverages language pseudo-labels extracted from a spoken language recognition model pre-trained on VoxLingua107 \cite{valk2021voxlingua107} dataset. To the best of our knowledge, we are the first to perform intra-domain disentangled representation learning using only pseudo-labels.

\section{Bilingual speaker recognition test set}
\label{sec:VoxCeleb1-B}

We publish a large-scale bilingual speaker recognition evaluation protocol derived from VoxCeleb1 dataset~\cite{nagrani2020voxceleb}, which is one of the widespread benchmark evaluation datasets in the recent speaker recognition field. Most of existing evaluation sets do not focus on the bilingual scenarios. Our speaker recognition evaluation set contains 808,574 trials in total. Half of the trials are intra-speaker cross-lingual and the remaining trials are inter-speaker monolingual.

\subsection{Obtaining language labels}

To simulate the bilingual scenarios with VoxCeleb1 dataset, it is necessary to have the language labels of the utterances in the test set. We utilise language annotations of VoxCeleb1 dataset from VoxSRC 2021~\cite{brown2022voxsrc} which are manually checked after obtaining language pseudo-labels of the utterances by using a Spoken Language Recognition (SLR) model pre-trained on VoxLingua107~\cite{valk2021voxlingua107} dataset. VoxLingua107 dataset contains 6,628 hours of speech that are divided into 107 languages.

Assuming that a single speaker speaks only one language in a video, one audio sample is randomly sampled for each video. 15 languages including English, French, Hindi, German, Spanish, Italian, Afrikaans, Portuguese, Dutch, Korean, Urdu, Swedish, Russian, Chinese, and Arabic are annotated by annotators of various nationalities. Out of 153,516 utterances in the VoxCeleb1 dataset, 883 utterances, whose language could not be recognised by annotators, have been excluded from the proposed evaluation list.

\subsection{{\tt VoxCeleb1-B} Evaluation list}

Speaker verification evaluation protocol consists of \textit{positive} and \textit{negative} trials.
Each trial involves an enrollment utterance and a test utterance. 
The trial type is decided based on whether the enrollment and the test utterances have the same speaker identity. 
To evaluate the robustness of speaker recognition models in the bilingual scenarios, we propose an evaluation protocol named VoxCeleb1-B, which simulates language-mismatch scenarios with a large amount of cross-lingual trials. Using the speaker and language annotations, we generate 404,287 intra-speaker cross-lingual trials and inter-speaker monolingual trials each. The number of speakers for each language and the number of samples per speaker are limited to 1,000 and 15, respectively, to avoid bias towards more frequent languages.

\Cref{tab:testsets} shows the statistics of existing evaluation lists derived from VoxCeleb1, and the proposed VoxCeleb1-B. 
The three original VoxCeleb1 test sets and the VoxSRC 2020~\cite{nagrani2020voxsrc} validation set are expected to contain very few cross-lingual positive trials, whereas the VoxSRC 2021~\cite{brown2022voxsrc} contains some cross-lingual trials. VoxCeleb1-B is explicitly designed to contain a large number of cross-lingual trials.



\section{Language-disentangled learning}
\label{sec:language-distentangled learning}

In this section, we describe the proposed language-disentangled representation learning strategy. Our training framework is inspired by~\cite{tzeng2017adversarial, chung2019delving, kang2022augmentation} and summarised in \Cref{fig:training_framework}.
The network consists of a speaker embedding network that includes a speaker feature extractor and a speaker embedding layer, a speaker classifier, and a language classifier. The speaker embedding network follows the existing speaker models~\cite{chung20b_interspeech, kwon2021ins} while the language classifier is attached for the purpose of a language discriminator.

The speaker embedding network produces frame-level embeddings ${z_i}$ from the input mel-spectrogram data $x_i \in \mathcal{R}^{{T}\times{F}}$ ($1 \leq i \leq N$), where $T$, $F$ and $N$ are the number of frames, frequency bins, and the size of mini-batch, respectively. To derive an utterance-level vector ${e_i}^{S}$ from the frame-level embeddings ${z_i}$, we adopt Attentive Pooling Layer (APL) which includes self-attentive pooling (SAP)~\cite{cai18_odyssey} or attentive statistics pooling (ASP)~\cite{okabe18_interspeech} as an speaker embedding layer.

The speaker embedding vector ${e_i}^{S}$ is passed as an input feature to both the speaker classifier and the language classifier, which consist of one and three fully-connected layers, respectively. We obtain the language feature vector, ${e_i}^{L}$, from the output of the second fully-connected layer in the language classifier. For the language classifier, the GRL is placed at the front of the language classifier and is activated in speaker embedding training step.

The training process of our framework alternates between two phases for the data from the same mini-batch: (1) language discriminator training, and (2) speaker embedding training. In the first phase, we train the language discriminator to recognise the spoken language from ${e_i}^{S}$. In the second phase, the speaker recognition network is trained to classify speakers, while intentionally trained to poorly recognise spoken languages.

\subsection{Language discriminator training}
In this step, we train the language classifier, while freezing the speaker recognition network. This approach can be interpreted to train the language recognition for the latest state of the speaker representation vector ${e_i}^S$ that has been extracted by the speaker embedding layer. The objective function $L_{lang}$ of the language classifier is a categorical cross-entropy loss. In~\Cref{fig:discriminator_training}, the parts of the network coloured in blue are optimised by $L_{lang}$.

\subsection{Speaker embedding training}
In this step, we train the speaker recognition network with language-disentangled representation learning. The language classifier's parameters are not updated at this stage. The total loss function to train the language-disentangled speaker recognition model can be formulated as follows.

\begin{equation}
    L_{total} = L_{spk} + L_{de}
\end{equation}

\noindent where $L_{spk}$ is an objective function for the speaker recognition and $L_{de}$ is an objective function of the disentangled representation learning. For $L_{spk}$, we can utilise objective functions such as softmax loss, prototypical loss, and contrastive loss, which have been employed in the previous works~\cite{chung20b_interspeech, kwon2021ins}. For prototypical loss and contrastive loss, we exclude the speaker classifier since these losses are directly derived from the speaker embedding vectors rather than speaker logits. For $L_{de}$, we can apply objective functions from two types of learning methods, namely adversarial learning-based method and metric learning-based method. In this work, we select gradient reversal layer as an adversarial learning-based method, and metric learning-based methods include cosine similarity minimisation and mean absolute Pearson's correlation minimisation. The details of each method are as follows.


\newpara{Gradient Reversal Layer (GRL).} Gradient reversal layer inverts the gradient value of target loss function to opposite sign for disturbing the convergence of the target loss function. In our work, the target loss function is $L_{lang}$.

\newpara{Cosine similarity (COS) minimisation.} This method minimises cosine similarity between speaker embedding vector ${e_i}^{S}$ and language feature vector ${e_i}^{L}$.

\newpara{Mean Absolute Pearson's Correlation (MAPC) minimisation.} This method minimises mean absolute Pearson's correlation~\cite{kang2020disentangled} between speaker embedding vector and language feature vector. In our work, $L_{corr}$ can be formulated as follows.

\newcommand{\Cov}{\operatorname{Cov}}

\begin{equation}
   \ L_{corr} = \frac{1}{N} \sum_{i=1}^{N}\sum_{j=1}^{F} \frac{|\Cov({e^S_{i,j}},{e^L_{i,j}})|}{\sigma({e^S_{i,j}})\cdot \sigma({e^L_{i,j}})}
\end{equation}

\noindent where $\Cov(\cdot)$ is the covariance and $\sigma(\cdot)$ is the standard deviation. ${F}$ is the dimensionality of the embedding vector $e_i$.

\newpara{Ours.} We propose an effective disentangled representation learning that consists of GRL and MAPC minimisation. The total loss function $L_{total}$ of our method can be formulated as follows.

\begin{equation}
    L_{total} = L_{spk} + L_{corr} + \lambda L_{lang}
\end{equation}

\noindent where $\lambda$ is a weight value for summation. In~\Cref{fig:representation_training}, the parts of the network coloured in yellow are optimised by $L_{total}$.







\section{Experiments}
\label{sec:experiments}

\begin{table*}
\centering
\caption{Equal Error Rates (EER) and minimum Detection Cost Function (minDCF) on (a) VoxSRC 2021 validation set, VoxCeleb1-B, (b) VoxCeleb1 test sets, and VoxSRC 2020 validation set. Accuracy of Spoken Language Recognition (SLR Acc.) is computed on VoxCeleb1-B. Lower SLR means less language information. 
All experiments except for spoken language recognition are repeated three times, and we report the mean and the standard deviation. {\bf GRL}: Gradient reversal layer; {\bf COS min.}: Cosine similarity minimisation; {\bf MAPC min.}: Mean absolute Pearson's correlation minimisation; {\bf Ours}: Combination of GRL and MAPC minimisation.}
\begin{subtable}[t]{0.70\textwidth}
\resizebox{1.0\linewidth}{!}{
\begin{tabular}[t]{@{\extracolsep{8pt}}lccccc@{}}
\toprule
\multirow{2}{*}{\bf Model} & \multicolumn{2}{c}{\bf VoxSRC 2021 Val} & \multicolumn{3}{c}{\bf VoxCeleb1-B}  \\
\cmidrule[0.2pt]{2-3} \cmidrule[0.2pt]{4-6}
& {\bf EER (\%)} & {\bf minDCF} & {\bf EER (\%)} & {\bf minDCF} & {\bf SLR Acc. (\%)} \\
\toprule
ResNet-S~\cite{chung20b_interspeech} &  9.22~$\pm$~0.15 & 0.503~$\pm$~0.007 & 9.69~$\pm$~0.14 & 0.617~$\pm$~0.007 & 87.2 \\
\hspace{3mm} {\it + GRL}  & 10.27~$\pm$~0.88 & 0.541~$\pm$~0.036 & 10.39~$\pm$~1.24 & 0.598~$\pm$~0.055 & 87.2 \\
\hspace{3mm} {\it + COS min.}  & 9.33~$\pm$~0.18 & 0.505~$\pm$~0.002 & 10.21~$\pm$~0.22 & 0.614~$\pm$~0.012 & 87.0 \\
\hspace{3mm} {\it + MAPC min.} & 8.85~$\pm$~0.12 & 0.486~$\pm$~0.005 & 9.85~$\pm$~0.15 & 0.594~$\pm$~0.011 & 86.8 \\
\hspace{3mm} Ours & {\bf 8.35~$\pm$~0.05} & {\bf 0.461~$\pm$~0.002} & {\bf 8.25~$\pm$~0.06} & {\bf 0.506~$\pm$~0.002} & 82.9 \\
\toprule
ResNet-L~\cite{kwon2021ins} & 5.16~$\pm$~0.08 & 0.308~$\pm$~0.010 & 5.96~$\pm$~0.23 & 0.397~$\pm$~0.016 & 88.3\\
\hspace{3mm} {\it + GRL} & 4.53~$\pm$~0.25 & 0.263~$\pm$~0.008 & 3.98~$\pm$~0.27 & 0.268~$\pm$~0.020 & 72.1 \\
\hspace{3mm} {\it + COS min.} & 5.21~$\pm$~0.18 & 0.317~$\pm$~0.015 & 5.99~$\pm$~0.35 & 0.423~$\pm$~0.016 & 88.8 \\
\hspace{3mm} {\it + MAPC min.} & 5.23~$\pm$~0.01 & 0.311~$\pm$~0.009 & 5.93~$\pm$~0.19 & 0.411~$\pm$~0.018 & 88.2 \\
\hspace{3mm} Ours & {\bf 4.22~$\pm$~0.03} & {\bf 0.246~$\pm$~0.006} & {\bf 3.69~$\pm$~0.12} & {\bf 0.254~$\pm$~0.010} & 80.1 \\
\bottomrule
\end{tabular}}
\caption{Results on VoxSRC 2021 validation set and VoxCeleb1-B.}
\label{tab:table2_a}
\end{subtable}


\begin{subtable}[t]{0.98\linewidth}
\resizebox{1.0\linewidth}{!}{
\begin{tabular}[t]{ @{\extracolsep{8pt}}lcccccccc@{}}

\toprule
\multirow{2}{*}{\bf Model} & \multicolumn{2}{c}{\bf VoxCeleb1 cl.} & \multicolumn{2}{c}{\bf VoxCeleb1-E cl.} & \multicolumn{2}{c}{\bf VoxCeleb1-H cl.} & \multicolumn{2}{c}{\bf VoxSRC 2020 Val}  \\
\cmidrule[0.2pt]{2-3} \cmidrule[0.2pt]{4-5} \cmidrule[0.2pt]{6-7} \cmidrule[0.2pt]{8-9}
& {\bf EER (\%)} & {\bf minDCF} & {\bf EER (\%)} & {\bf minDCF} & {\bf EER (\%)} & {\bf minDCF} & {\bf EER (\%)} & {\bf minDCF} \\
\toprule
ResNet-S &  2.24~$\pm$~0.13 & 0.174~$\pm$~0.005 & 2.43~$\pm$~0.04 & 0.175~$\pm$~0.003 & 4.74~$\pm$~0.07 & 0.299~$\pm$~0.005 & 6.91~$\pm$~0.07 & 0.393~$\pm$~0.004  \\
\hspace{3mm} {\it + GRL}  & 2.98~$\pm$~0.14  & 0.210~$\pm$~0.012 & 3.12~$\pm$~0.19 & 0.222~$\pm$~0.013 & 5.72~$\pm$~0.37 & 0.351~$\pm$~0.020 & 8.07~$\pm$~0.47 & 0.457~$\pm$~0.025  \\
\hspace{3mm} {\it + COS min.}  & {\bf 2.13~$\pm$~0.08} & 0.164~$\pm$~0.006 & 2.45~$\pm$~0.06 & 0.178~$\pm$~0.004 & 4.78~$\pm$~0.12 & 0.303~$\pm$~0.006 & 6.83~$\pm$~0.06 & 0.388~$\pm$~0.004  \\
\hspace{3mm} {\it + MAPC min.} & 2.16~$\pm$~0.07  & {\bf 0.157~$\pm$~0.003} & {\bf 2.33~$\pm$~0.01} & {\bf 0.166~$\pm$~0.001} & {\bf 4.48~$\pm$~0.02} & {\bf 0.284~$\pm$~0.003} & 6.61~$\pm$~0.04 & {\bf 0.373~$\pm$~0.003}  \\
\hspace{3mm} Ours  & 2.15~$\pm$~0.01 & 0.172~$\pm$~0.001 & 2.42~$\pm$~0.01 & 0.171~$\pm$~0.000 & 4.49~$\pm$~0.02 & {\bf 0.284~$\pm$~0.001} & {\bf 6.54~$\pm$~0.02} & 0.378~$\pm$~0.001 \\
\toprule
ResNet-L &  1.17~$\pm$~0.00  & 0.083~$\pm$~0.003 & 1.30~$\pm$~0.01 & 0.091~$\pm$~0.001 & 2.58~$\pm$~0.02 & 0.164~$\pm$~0.001 & 4.06~$\pm$~0.02 & 0.231~$\pm$~0.005 \\
\hspace{3mm} {\it + GRL} & 1.22~$\pm$~0.04 & 0.088~$\pm$~0.009 & 1.34~$\pm$~0.04 & 0.096~$\pm$~0.002 & 2.58~$\pm$~0.02 & 0.166~$\pm$~0.001 & 4.13~$\pm$~0.04 & 0.230~$\pm$~0.003 \\
\hspace{3mm} {\it + COS min.}  & 1.11~$\pm$~0.01 & 0.084$\pm$~0.007 & 1.25~$\pm$~0.03 & 0.091~$\pm$~0.004 & 2.52~$\pm$~0.02 & 0.164~$\pm$~0.002 & 3.99~$\pm$~0.03 & 0.228~$\pm$~0.002  \\
\hspace{3mm} {\it + MAPC min.} & 1.10~$\pm$~0.02 & {\bf 0.079~$\pm$~0.001} & {\bf 1.24~$\pm$~0.02} & {\bf 0.088~$\pm$~0.002} & 2.48~$\pm$~0.04 & 0.160~$\pm$~0.002 & 3.99~$\pm$~0.04 & 0.224~$\pm$~0.001  \\
\hspace{3mm} Ours & {\bf 0.99~$\pm$~0.05} & {\bf 0.079~$\pm$~0.004} & 1.25~$\pm$~0.01 & {\bf 0.088~$\pm$~0.000} & {\bf 2.42~$\pm$~0.04} & {\bf 0.154~$\pm$~0.003} & {\bf 3.91~$\pm$~0.06} & {\bf 0.220~$\pm$~0.001}  \\
\bottomrule
\end{tabular}}
\caption{Results on cleaned version of VoxCeleb1 test sets and VoxSRC 2020 validation set.}
\label{tab:table2_b}
\end{subtable}

\vspace{-3mm}
\label{tab:results_main}
\end{table*}

\subsection{Input representations and model architecture}
For the input representation of the neural network, we use log-mel spectrograms that are extracted with a hamming window, 25ms window size and 10ms stride size.

We focus on demonstrating the effectiveness of the proposed learning strategy and its compatibility with previous models. Thus, we employ two existing variants~\cite{chung20b_interspeech, kwon2021ins} of the 34-layer residual network, and rename each variant as ResNet-S~\cite{chung20b_interspeech} and ResNet-L~\cite{kwon2021ins}. ResNet-S uses the SAP~\cite{cai18_odyssey} and the angular prototypical loss, and ResNet-L uses the ASP~\cite{okabe18_interspeech} and the angular prototypical loss combined with the softmax loss, which is in line with the original papers. The output size of each classifier is equal to the number of each task's classes.

\subsection{Disentangled representation learning method}
We evaluate various disentangled representation learning strategies in terms of separating irrelevant information from intra-domain speaker representation rather than domain adaptation in cross-domain. We perform extensive experiments on various disentangled representation learning strategies including adversarial learning-based method using the GRL, metric learning-based method using COS minimisation or MAPC minimisation, and the proposed method, which is the combination of GRL and MAPC minimisation.

\subsection{Implementation details}
\newpara{Datasets.}
We use the development partition with 5,994 speakers of the VoxCeleb2 \cite{chung18b_interspeech} as the training dataset. In order to learn language-disentangled representation, we use the language pseudo-labels of VoxCeleb2 dataset extracted from an SLR model pre-trained on VoxLingua107 dataset. For evaluation, we use the three original test sets based on VoxCeleb1~\cite{nagrani2020voxceleb}, the VoxSRC validation sets~\cite{nagrani2020voxsrc, brown2022voxsrc} and VoxCeleb1-B, which is the proposed large-scale bilingual speaking evaluation set.

\newpara{Training.}
Our implementation is based on the PyTorch framework~\cite{paszke2019pytorch}. We use the Adam Optimizer~\cite{kingma20153rd} with initial learning rate of 0.001 decreasing by $3\%$ every epoch. All experiments are performed on a single NVIDIA A5000 GPU with 24GB memory. We use the batch size of 500 and 300 for ResNet-S and ResNet-L, respectively. The training takes around 3 days. The $\lambda$ value is set to 0.5.

\subsection{Evaluation protocol}
We report Equal Error Rate (EER)  where the False Rejection Rate (FRR) and the False Alarm Rate (FAR) are equal, and minimum Detection Cost Function (minDCF)~\cite{omid19nsre} which is a weighted sum of FRR and FAR. For each trial, we sample each utterance into ten 4-second segments and compute similarities between all possible combinations of segment pairs. We use the mean of the similarities as a score of the trial.  This evaluation protocol is in line with that from~\cite{chung18b_interspeech, chung2019delving, chung20b_interspeech, kang2022augmentation}.

\section{Results}
The experimental results are summarised in~\Cref{tab:results_main}. Specifically, \Cref{tab:table2_a} reports the results on the test sets that mainly consider bilingual scenario, while \Cref{tab:table2_b} contains the results on existing test sets that do not take the bilingual speakers into account during their construction phases. We train all models 3 times with different random seeds, and report the mean and the standard deviation.

\newpara{Bilingual scenario in speaker recognition.} As shown in~\Cref{tab:table2_a}, most of the models show poor performance on the VoxSRC 2021 validation set and VoxCeleb1-B. In particular, all baselines show the lowest performance on VoxCeleb1-B which is composed entirely of bilingual trials. This implies that bilingual scenario is one of the most demanding challenges in the speaker recognition field.

\newpara{Intra-domain disentangled representation learning.}
When compared to the baseline of ResNet-L, the use of GRL shows notable performance improvements of $33\%$ and $12\%$ in VoxCeleb1-B and VoxSRC2021 validation set, respectively, while the performance degrades on the other evaluation sets. Furthermore, ResNet-S with GRL exhibits the lowest performance and the highest standard deviation on every evaluation set including VoxCeleb1-B. This highlights the drawback of GRL, which tends to induce unstable training despite its effectiveness in removing the language information from the speaker representation.

We observe out that the metric learning-based methods, COS and MAPC minimisation, outperform the baselines on VoxCeleb1 test sets and VoxSRC 2020 validation set. However, we observe no performance improvement on VoxCeleb1-B. This suggests that the disentangled representation learning method which utilises metric learning performs a role of regularisation, but fails to isolate the language information from the speaker representation. As a result, we verify that existing disentangled representation learning strategies applied to cross-domain tasks do not guarantee generalisation to intra-domain tasks such as bilingual scenarios.

The proposed learning method on ResNet-L outperforms the baselines and all existing methods on most of evaluation sets except VoxCeleb1-E test set. In the case of ResNet-S, our method shows the best performance on VoxSRC validation sets and VoxCeleb1-B while remaining effective on the other evaluation sets. Especially, ResNet-S and ResNet-L exhibit significant performance improvements by 15\% and 38\% on VoxCeleb1-B, respectively. This demonstrates that the proposed learning strategy overcomes the limitations of existing disentangled representation learning methods and facilitates robust language-disentangled speaker representation learning.

\newpara{The use of pseudo-labels.} Training of all experiments are performed with language pseudo-labels of VoxCeleb2. Nonetheless, the proposed learning strategy works successfully, showing significant performance improvements in bilingual scenarios. This indicates that it can be cost-effective to use pseudo-labels of specific factor in speaker recognition, which can be extended to other factors of variation that must be disentangled from speaker embeddings.

\newpara{Language-disentangled speaker representation.} To verify whether the language information is separated from the speaker representation, we evaluate a spoken language recognition model trained from scratch with the speaker embedding vector extracted from each model as input data. The structure of the spoken language recognition model is the same as the language classifier described in Section \ref{sec:language-distentangled learning}. As shown in~\Cref{tab:table2_a} (SLR Acc.), our method shows lower spoken language recognition performance than the baselines of two models. This highlights that the proposed method successfully removes the language information from the speaker representation.

\section{Conclusion}
\label{sec:conclusion}
We have developed strategies to train speaker embeddings that are robust to bilingual speaking scenarios, and proposed an evaluation protocol that takes bilingual speakers into account.
Our large-scale evaluation protocol is designed to analyse speaker recognition performance under bilingual scenarios, and we make this evaluation set publicly available. 
We also propose a new learning strategy to resolve the bilingual problem. Our learning strategy disentangles language information from the speaker representation in order to make the embeddings robust to cross-lingual trials. 
Our proposed learning strategy shows significant performance improvements under bilingual scenarios, while remaining effective on existing test sets.
\section{Acknowledgements}
\label{sec:acknowledgements}
We would like to thank Icksang Han and Bong-Jin Lee for helpful comments.

\clearpage
\bibliographystyle{IEEEtran}
\bibliography{shortstrings,mybib}

\end{document}